\documentclass[a4paper,useAMS,usenatbib]{mnras}
\usepackage{graphics,epsfig,psfig}
\hypersetup{colorlinks=true,allcolors=blue}
\usepackage{todonotes}
\usepackage[normalem]{ulem}
\usepackage{xcolor}
\usepackage{float}
\usepackage{booktabs}
\usepackage[style=base, singlelinecheck=off, font=small, labelfont=bf, 
justification=raggedright]{caption}
\usepackage{subfigure}
\usepackage{orcidlink}

\usepackage{graphicx}
\usepackage[]{inputenc,amssymb}
\usepackage{natbib}
\usepackage{url}
\usepackage{widetext}
\usepackage{amsmath}
\usepackage{cancel}

  \makeatletter
\def\@fnsymbol#1{\ensuremath{\ifcase#1\or \dagger\or \ddagger\or
   \mathsection\or \mathparagraph\or \|\or **\or \dagger\dagger
   \or \ddagger\ddagger \else\@ctrerr\fi}}
    \makeatother
\def \be{\begin{equation}}
\def \ee{\end{equation}}
\def \bea{\begin{eqnarray}}
\def \eea{\end{eqnarray}}

\def\apj{ApJ}
\def\mnras{MNRAS}
\def\prd{PRD}
\def\aap{A\&A}
\definecolor{webgreen}{rgb}{0,.5,0}
\definecolor{webbrown}{rgb}{.6,0,0}

\def\aap{A\&A}
\def\apj{ApJ}
\def\apjs{ApJS}

\def\mnras{MNRAS}

\def\aj{AJ}
\def\nat{Nature}

\def\prd{Phys.Rev.D}         
\def\pasp{PASP}              

\title[Synergy between \texttt{LVK}, \texttt{DESI} and \texttt{SPHEREx}]{Mapping the cosmic expansion history from LIGO-Virgo-KAGRA in synergy with DESI and SPHEREx}
\author[Cigarrán Díaz and Mukherjee (2021)]{Cristina Cigarrán Díaz$^{1}$, Suvodip Mukherjee$^{1,2,3,4}$\thanks{Corresponding author (s.mukherjee@uva.nl)}\thanks{The author list is in the alphabetical order.}\orcidlink{0000-0002-3373-5236}
\\
$^{1}$ Gravitation Astroparticle Physics Amsterdam (GRAPPA),
Anton Pannekoek Institute for Astronomy and Institute for Physics,\\
University of Amsterdam, Science Park 904, 1090 GL Amsterdam, The Netherlands\\
$^2$ Institute Lorentz, Leiden University, PO Box 9506, Leiden 2300 RA, The Netherlands\\
$^3$Delta Institute for Theoretical Physics, Science Park 904, 1090 GL Amsterdam, The Netherlands\\
$^4$Perimeter Institute for Theoretical Physics, 31 Caroline Street N., Waterloo, Ontario, N2L 2Y5, Canada\\}
\pubyear{2021}
\begin{document}
\label{firstpage}
\pagerange{\pageref{firstpage}--\pageref{lastpage}}
\maketitle

\label{firstpage}

\begin{abstract}
The measurement of the expansion history of the Universe from the redshift unknown gravitational wave (GW) sources (dark GW sources) detectable from the network of \texttt{LIGO-Virgo-KAGRA} (\texttt{LVK}) detectors depends on the synergy with the galaxy surveys having accurate redshift measurements over a broad redshift range,  large sky coverage, and detectability of fainter galaxies. In this work, we explore the possible synergy of the LVK with the spectroscopic galaxy surveys such as \texttt{DESI} and \texttt{SPHEREx} to measure the cosmological parameters which are related to the cosmic expansion history and the GW bias parameters. We show that by using the three-dimensional spatial cross-correlation between the dark GW sources and the spectroscopic galaxy samples, we can measure the value of Hubble constant with about $2\%$ and $1.5\%$ precision from \texttt{LVK+DESI} and \texttt{LVK+SPHEREx} respectively within the five years of observation time with $50\%$ duty-cycle. Similarly, the dark energy equation of state can be measured with about $10\%$ and $8\%$ precision from \texttt{LVK+DESI} and \texttt{LVK+SPHEREx} respectively.  We find that due to the large sky coverage of \texttt{SPHEREx} than \texttt{DESI},  performance in constraining the cosmological parameters is better from the former than the latter. By combining \texttt{Euclid} along with \texttt{DESI} and \texttt{SPHEREx}, a marginal gain in the measurability of the cosmological parameters is possible from the sources at high redshift ($z\geq 0.9$). 
\end{abstract}

\section{Introduction}
Mapping the cosmic expansion history at different redshift provides a direct way to unveil the constituents of the Universe. Gravitational waves (GWs) provide a robust way to achieve this with the aid of coalescing binary compact objects such as binary neutron stars (BNSs), neutron star black holes (NSBHs), and binary black holes (BBHs).  These are \textit{standard sirens} if the emission of GWs from these systems can be completely predictable by the General Theory of Relativity and redshift to these sources can be measured independently, as shown for the first time in the seminal work by \cite{Schutz}. Though the direct measurement of the source redshift using the GW signal is not possible due to the mass-redshift degeneracy, several techniques are proposed such as the spectroscopic measurement of the redshift of the host galaxy from electromagnetic counterpart \citep{Schutz, Holz:2005df, Dalal:2006qt, PhysRevD.77.043512, 2010ApJ...725..496N, 2011CQGra..28l5023S, Chen:2017rfc,Feeney:2018mkj}, tidal-deformability of neutron star \citep{Messenger:2011gi}, known mass-scale \citep{Farr:2019twy, Mastrogiovanni:2021wsd}, statistical host identification \citep{DelPozzo:2012zz,Arabsalmani:2013bj, Chen:2017rfc, Nair:2018ign, PhysRevD.101.122001, Fishbach:2018gjp, Abbott:2019yzh, Soares-Santos:2019irc, Abbott:2020khf}, and exploring the spatial clustering-scale between the GW sources and galaxies \citep{PhysRevD.93.083511,Mukherjee:2018ebj, Mukherjee:2019wcg, Mukherjee:2020hyn, Bera:2020jhx,Mukherjee:2020mha,2021ApJ...918...20C} to independently infer the redshift of the GW sources. With the aid of these techniques, it is expected that using the plethora of GW sources which are detectable in the coming years from the current generation GW detectors such as \texttt{Laser Interferometer Gravitational-wave Observatory (LIGO)} \citep{LIGOScientific:2014pky,Martynov:2016fzi}, \texttt{Virgo} \citep{Acernese_2014,PhysRevLett.123.231108}, \texttt{Kamioka Gravitational Wave Detector (KAGRA)} \citep{Akutsu:2018axf,KAGRA:2020tym}, \texttt{LIGO-India} \citep{Unnikrishnan:2013qwa}\footnote{\href{https://dcc.ligo.org/LIGO-M1100296/public }{LIGO-India, Proposal of the Consortium for Indian Initiative in Gravitational-wave Observations (IndIGO).}}, we can achieve accurate and precise measurement of the expansion history, if detector systematics can be mitigated successfully \citep{Sun:2020wke, Bhattacharjee:2020yxe, Vitale:2020gvb}. 

Such an independent measurement is going to play a vital role in resolving the tension in the value of the current expansion rate of the Universe known as the Hubble constant between the late Universe probes \citep{Riess:2019cxk,Riess:2019qba,Wong:2019kwg,Freedman:2019jwv,Freedman:2020dne,Soltis:2020gpl} and the early Universe probes \citep{Spergel:2003cb, Spergel:2006hy,2011ApJS..192...18K, Ade:2013zuv, Ade:2015xua,Alam:2016hwk, Aghanim:2018eyx,eBOSS:2020yzd,ACT:2020gnv} and will also be capable to shed light on the nature of dark matter and dark energy which constitutes  about $95\%$ of the energy budget of the Universe, as indicated by the measurements from electromagnetic  probes such as supernovae \citep{Perlmutter:1998np,2009ApJ...695..287R, Riess:2019cxk}, cosmic microwave background \citep{Spergel:2003cb, Spergel:2006hy,2011ApJS..192...18K, Ade:2013zuv, Ade:2015xua, Aghanim:2018eyx,ACT:2020gnv}, and baryon acoustic oscillation \citep{Eisenstein:1997ik,Eisenstein:1997jh,2013AJ....145...10D,Alam:2016hwk,eBOSS:2020yzd}. 

Since the discovery of GW \citep{PhysRevLett.116.061102} by the \texttt{LIGO-Virgo Scientific Collaboration} and followed by the detection of about 50 GW sources until the first half of the third observation run \citep{Abbott_2019, Abbott_2021}, we have seen one BNS event GW170817 \citep{TheLIGOScientific:2017qsa} with electromagnetic counterpart \citep{Kasliwal:2017ngb, GBM:2017lvd}, which enabled the first measurement of Hubble constant $H_0= 70_{-8}^{+12}$ km/s/Mpc. A revised value after incorporating with a different peculiar velocity correction scheme is obtained by \citet{Howlett:2019mdh, Mukherjee:2019qmm, Nicolaou:2019cip}.  For one of the BBH event GW190521 \citep{PhysRevLett.125.101102} detected by the \texttt{LIGO-Virgo} detectors, there was a plausible electromagnetic counterpart detected by the \texttt{Zwicky transient facility (ZTF)} \citep{PhysRevLett.124.251102}, which provided only a weak measurement of the value of Hubble constant $H_0=  62.2_{-19.7}^{+29.5}$ km/s/Mpc \citep{Mukherjee:2020kki} for the SEOBNRv4PHM waveform \footnote{$H_0=  50.4_{-19.5}^{+28.1}$ km/s/Mpc for the NRSur7dq4 waveform, and $H_0= 43.1_{-11.4}^{+24.6}$ km/s/Mpc for the IMRPhenomPv3HM waveform.}.  Measurement of the value of Hubble constant using GW190521 is also performed by several other independent studies \citep{Chen:2020gek, Gayathri:2021isv, Mastrogiovanni:2020mvm}. The sources which did not show any electromagnetic counterpart (also called dark sirens), were used to measure the value of the Hubble constant by the statistical host identification technique \citep{PhysRevD.101.122001} using the sources detected in the first two observation runs (O1+O2) \citep{Abbott:2019yzh} using the \texttt{GLADE} catalog \citep{Dalya:2018cnd}. More recently, using the GW events from O1+O2+O3a, a value of $H_0=67.3_{-17.9}^{+27.6}$ km/s/Mpc is obtained by \citet{Finke:2021aom}.  

The discovery made by the \texttt{LIGO-Virgo} detectors agrees with the current understanding that we only expect electromagnetic counterparts from GW sources such as BNS and NS-BH systems \citep{Foucart:2018rjc}. For BBHs, unless there is a presence of baryonic matter \citep{McKernan:2019hqs, PhysRevLett.123.181101, Tagawa:2020jnc}, electromagnetic counterparts are not expected in standard scenarios. However, the number of BBHs is expected to be more as it can be detected up to a large luminosity distance and hence can encompass more cosmic volume. So to use these large detectable BBHs as standard sirens to study cosmology, robust techniques are required to make an accurate and precise measurement of the redshift is required. 

One such method that is free from any astrophysical assumption about the GW sources (or population) is the cross-correlation technique \citep{Mukherjee:2018ebj, Mukherjee:2019wcg, Mukherjee:2020hyn}. This technique explores the spatial clustering between galaxies and GW sources to infer the clustering redshift of the GW sources.  {The success of this method depends on the availability of spectroscopic galaxy surveys over a large sky area and redshift range.} For a single galaxy survey to satisfy both criteria is difficult to achieve. But by combining multiple galaxy surveys, one can reach the requirements to apply the cross-correlation technique to dark standard sirens. 

In this work, we explore  {for the first time} the synergy between the upcoming spectroscopic galaxy surveys such \texttt{Dark Energy Spectroscopic Instrument} (\texttt{DESI}) \citep{desi} and \texttt{Spectro-Photometer for the History of the Universe, Epoch of Reionization, and Ices Explorer} (\texttt{SPHEREx}) \citep{spherex} with the GW sources which are detectable from the network of four GW detectors (\texttt{LIGO-Livingston}, \texttt{LIGO-Hanford}, \texttt{Virgo}, and \texttt{KAGRA}), hereafter \texttt{LVK} \citep{KAGRA:2013pob}, for different scenarios of GW merger rates motivated from the \citet{Madau2014} star formation rate. We show the expected overlap between GW merger rates and galaxy surveys for different delay time scenarios and apply the cross-correlation technique on the simulated GW sources for the network of four detectors using the publicly available package \texttt{Bilby} \citep{bilby} with the mock \texttt{DESI} and  \texttt{SPHEREx} galaxy samples. 
 {This is the first work that has explored the implementation of the cross-correlation technique with the upcoming galaxy surveys taking into account the uncertainties in the GW sector and galaxy sector in a Bayesian framework. This work will help in initiating joint studies between \texttt{LVK} and upcoming galaxy surveys such as \texttt{DESI} and  \texttt{SPHEREx} to study cosmology.}
This paper is organized as follows, in Sec. \ref{gwsource}, we discuss the GW source population and merger rates for different delay time scenarios. In Sec. \ref{basic}, we set the detail formalism for the cross-correlation method between GW sources and galaxies.  In Sec. \ref{synergy} and  Sec. \ref{surveys} we will discuss the requirements for exploring the synergies between galaxy surveys and dark standard sirens and brief descriptions of the upcoming galaxy surveys respectively.  {In Sec. \ref{bayes} we discuss the Bayesian framework for estimating the cosmological parameters using the cross-correlation technique.} In Sec. \ref{mock} we discuss the simulated mock GW and galaxy catalog considered in this analysis and Sec. \ref{results} we make forecasts of estimating the cosmological parameters from \texttt{LVK}+\texttt{DESI} and \texttt{LVK}+\texttt{SPHEREx} combinations with the five years of \texttt{LVK} observation run in its design sensitivity with $50\%$ duty cycle. We conclude in Sec. \ref{conclusion} with the future scope of this study.

\section{GW Source properties}\label{gwsource}
\subsection{GW mass distribution}

The GWTC-2 \citep{Abbott_2021} catalog includes the first half of the third observing run, O3a \citep{Abbott_2021}, and the first and second observing run, O1 and O2 \citep{Abbott_2019}, of the \texttt{Advanced LIGO} \citep{aLIGO} and \texttt{Advanced Virgo} \citep{Acernese_2014} gravitational wave detectors. The GWTC-2 includes 44 new BBH events over GWTC-1 and thus allows for the study of the properties of the BBHs population, \citep{Abbott2020b}. In \citet{Abbott2020b}, several models are proposed to study the mass distribution of GW events from BBHs such as the Truncated model, Power-law + Peak model, Broken power-law, and Multi peak model. The Bayes factor estimation from GWTC-2 indicates a better fit for the Power-law + Peak model over the other models \citep{Abbott_2021}. However, we still are at an early stage to draw any conclusion on the mass distribution of the binary black holes and compare with the possible formation scenarios \citep{Coleman_Miller_2002, Gultekin_2004, Antonini_2016, PhysRevD.100.043027}.

\subsection{GW merger rate models}
\label{merger_rate}

With the observation of the BBH merger GW150914, \citep{AbbottGW150914}, we learned that BBHs indeed merge within the age of the Universe at a detectable rate. Astrophysical black holes (ABHs) and other astrophysical compact objects are produced from the death of stars. So, they are likely to be related to the star formation rate of the Universe. However, the delay time $t_d^{eff}$ between the formation and merger of compact objects is still uncertain, it may range between a few 100 Myr to around the age of the Universe \citep{2010ApJ...716..615O,2010MNRAS.402..371B, 2012ApJ...759...52D, Dominik:2014yma, 2016MNRAS.458.2634M, Lamberts:2016txh, 2018MNRAS.474.4997C, Elbert:2017sbr, Eldridge:2018nop, Buisson:2020hoq,Santoliquido:2020axb}. 

\begin{figure}
    \centering
    \includegraphics[trim={0.cm 0.cm 0.cm 0.cm},clip,width=1.\linewidth]{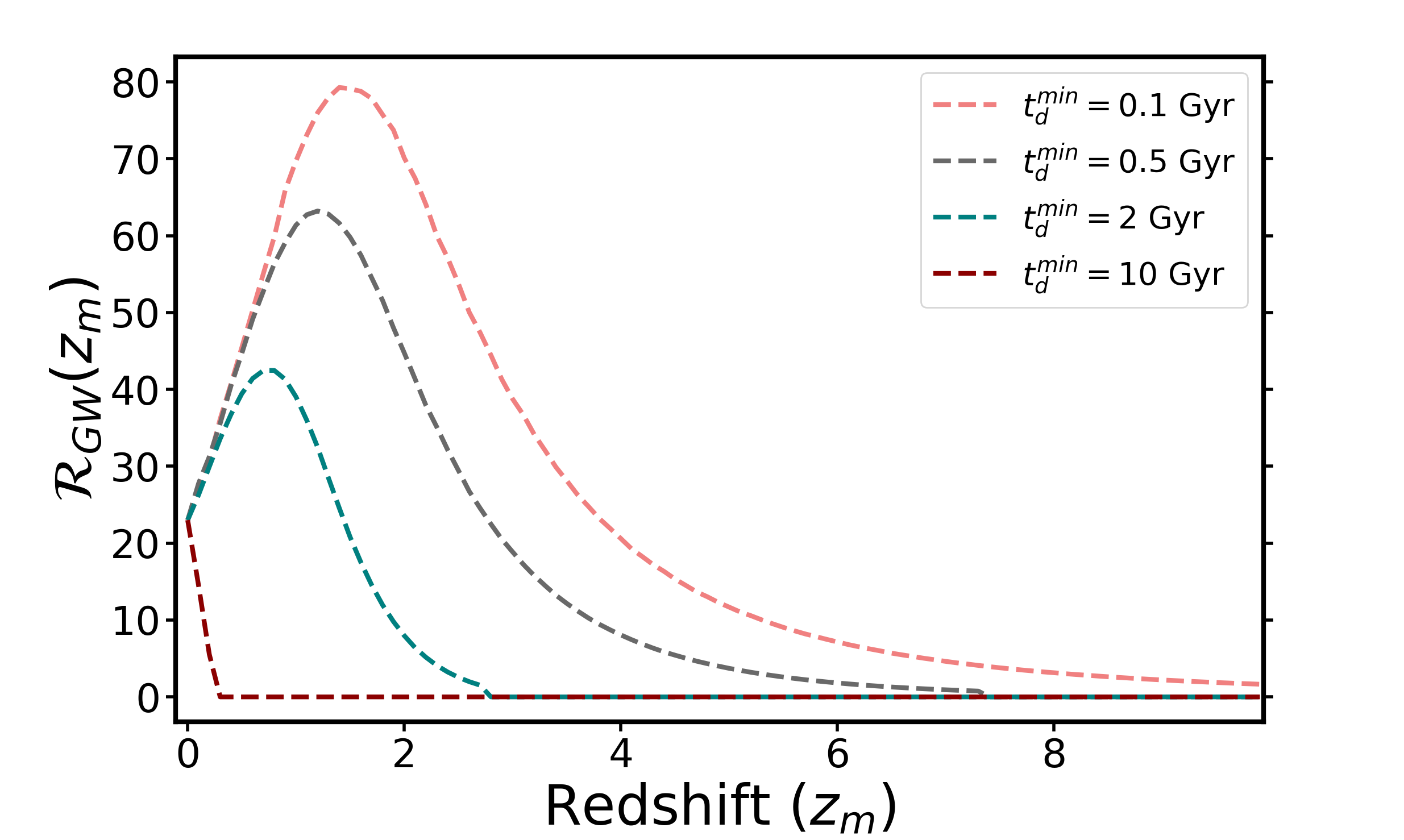}
    \caption{Merger rate of the GW sources for power-law delay time models, for different values of $t_d^{min}$ (in Gyr). The merger rate is higher for higher redshifts when the value of the delay time is small, while if the value of the delay time is larger the merger rate does not reach such high values at higher redshifts. Furthermore, we can see how the peak of the distribution moves to lower redshifts as we consider higher delay time values.}
    \label{fig:plot_powerlaw}
\end{figure}

The GW merger rate model we consider in this analysis is motivated by the star formation rate model well described by the \citet{Madau2014} relation

\begin{equation}\label{mdsfr}
R_{SFR}(z)= 0.015 \frac{(1+z)^{2.7}}{1+ (\frac{1+z}{2.9})^{5.6}} \mathrm{M}_\odot \; \mathrm{Mpc}^{-3} \; \mathrm{yr}^{-1}.
\end{equation}
This SFR model is based on galaxy surveys having  rest-frame FUV (far ultra-violet) and FIR (far infrared) measurements. This rate peaks around redshift of a few, $z \sim 2$ \citep{Madau2014}. The GW source merger rate  {at a redshift $z_m$} can then be written as \citep{2010ApJ...716..615O,2010MNRAS.402..371B, 2012ApJ...759...52D, Dominik:2014yma, 2016MNRAS.458.2634M, Lamberts:2016txh, 2018MNRAS.474.4997C, Elbert:2017sbr, Eldridge:2018nop, Vitale:2018yhm, Buisson:2020hoq,Santoliquido:2020axb} 
\begin{equation}\label{a1}
    \mathcal{R}_{GW}(z_m)= \mathcal{N}\int^{\infty}_{z_m} dz \frac{dt^{eff}_d}{dz}P(t^{eff}_d) R_{SFR}(z),
\end{equation}
where $\mathcal{N}$ is the normalization such that $\mathcal{R}_{GW}(z=0)= 25$ Gpc$^{-3}$ yr$^{-1}$ so it is in agreement with observations from GWTC-2, \citep{Abbott_2020},  and $t_d^{eff}$ is the effective delay time parameter. $R_{SFR}(z)$ is the star formation rate in Eq. \ref{mdsfr}. The effective delay time parameter $t_d^{eff}$ probability distribution function is not yet known from observation and thus needs to be modelled. We consider a power-law model for the delay time distribution $P(t_d^{eff})$ in our analysis \citep{Mukherjee:2021ags,Mukherjee:2021qam}.

\begin{equation}
        P(t_d^{eff}) = 1/t_d^{eff}, \quad t^{eff}_d > t^{min}_d,
\end{equation}
where $t_d^{min}$ in the minimum value of the effective delay time.   {For the BBHs to merge at redshift $z=0$, the maximum delay time is considered as the age of the Universe for the Planck cosmology \citep{Aghanim:2018eyx}.} We show in Fig. \ref{fig:plot_powerlaw} the GW merger rates for different values of the delay time parameter $t_d^{eff}$\footnote{ {The delay time for individual binary objects can depend on the property of the host galaxy, the mass of the GW sources, and their formation channels. As the actual observed delay time will be a combination of different delay times, we define a single delay time parameter and call it an effective delay time in this paper.}  }. The longer delay time causes more shifts in the peak of the GW merger rate to a lower redshift than the star formation peak $z \sim 2$. As a result, if the value of the delay time is larger than a few Gyrs, then most of the GW mergers will be happening at a lower redshift and the corresponding galaxy survey needs to have a window function overlapping with the merger rate of the GW sources. We discuss this in further detail in the later section of the paper. 

\section{Basic formalism of cross-correlation between GW sources and galaxies}\label{basic}

The three-dimensional cross-correlation technique for GW sources and galaxies is developed by \citet{Mukherjee:2020hyn,Mukherjee:2020mha}. The exploration of the spatial clustering between the GW sources and galaxies can be useful to measure the clustering redshift \citep{Newman:2008mb, Menard:2013aaa, Schmidt:2013sba} of the GW sources. We can write the galaxy density field as
\begin{equation}
    \delta_g (\mathbf{r}) = \frac{n_g(\mathbf{r})}{\bar{n}_g} - 1, 
\end{equation}
where $n_g(\mathbf{r})$ is the number density of galaxies at a position $\mathbf{r}$ and $\bar{n}_g$ is the mean number density of galaxies. We can relate the spatial distribution of galaxies with the underlying dark matter distribution in the Fourier space $\delta_m (\mathbf{k})$ as
\begin{equation}
    \delta_g(\mathbf{k},z) = b_g(k,z)\delta_m(\mathbf{k},z), 
\end{equation}
where $b_g(k,z)$ is the galaxy bias parameter which can be both scale and redshift dependent. 

Along with the Hubble flow, galaxies also have intrinsic  peculiar velocities that will displace them along the line of sight in redshift space and these displacements causes redshift-space distortions (RSD) and its signature gets imprinted in the clustering of galaxies in redshift space. The amplitude of distortions on large scales gives a measure of the linear RSD parameter $\beta_g \equiv f/b_g(k,z)$, which defined in terms of $f \equiv d\; \mathrm{ln}\; D / d \; \mathrm{ln}\; a$, with $D$  being the growth function, and $\mu_{\hat{k}} = \hat{n}\cdot \hat{k}$ is the cosine of the angle between the line of sight and the Fourier mode $\hat{k}$ \citep{Hamilton_1998, 1987MNRAS.227....1K}. The growth factor $D$ is defined as $D = \frac{H(z)}{H_0} \int_{z}^{\infty} \frac{dz' (1+z')}{H(z')^3} \left[ \int_{0}^{\infty} \frac{dz'' (1+z'')}{H(z'')^3} \right]{-1}$, and it measures the growth of the cosmological perturbations with redshift. The galaxy density field in redshift space $\delta^s_g (\mathbf{k},z)$ can be written in terms of the physical space density field $\delta_m^r(\mathbf{k},z)$ by including the effect from RSD as
\begin{equation}
    \delta^s_g (\mathbf{k},z)=b_g(k,z)(1+\beta_g \mu_{\hat{k}}) \delta_m^r(\mathbf{k},z).
\end{equation}

GW events from astrophysical sources will take place in galaxies and will therefore follow the spatial distribution of galaxies with a bias parameter $b_{GW}(k,z)$ - the value of this parameter is different from the galaxy bias parameter $b_g(k,z)$. We can then define the gravitational wave density field in real space as
\begin{equation}
    \delta_{GW}(\mathbf{k}, z) = b_{GW} (k,z)\delta_m^r(\mathbf{k},z),
\end{equation}
where $b_{GW} (k,z)$ is the gravitational wave bias parameter which measures how GW sources trace the large scale structure of the Universe \citep{Mukherjee:2019oma, Mukherjee:2020hyn}. There are also other studies to explore the GW bias parameter \citep{Scelfo:2018sny,Calore_2020, vijaykumar2020probing,Scelfo:2020jyw}.  {The RSD equivalent term is not included in the GW sector because of the error on the luminosity distance is large. Any additional anisotropic structure due to the line of sight velocity field will be sub-dominant in comparison to the luminosity distance error achievable from LVK. For the third generation, GW detectors such as Cosmic Explorer  \citep{Reitze:2019iox,2019CQGra..36v5002H}, and Einstein Telescope \citep{Punturo:2010zz}, contributions from velocity field and also weak lensing can be important. As we only focus on the LVK detectors in this work, we do not include RSD equivalent effects on GW luminosity distance.}

The clustering in the matter distribution can be described by the correlation function $\xi(r)$ which is related to the power spectrum by Fourier transformation \citep{1975ApJ...196....1P, 1977ApJ...216..665M, 1977ApJ...217..331M,  1977ApJS...34..425D, 1983ApJ...267..465D, Hamilton_1998, 1993ApJ...412...64L}. The spatial clustering of galaxies and GW can be written in Fourier space in terms of the auto-power spectrum and cross power-spectrum at different values of the redshift as \citep{Mukherjee:2020hyn}\footnote{ The notation $\langle . \rangle$ denotes the ensemble average.}

\begin{widetext}
\begin{align}
 & \left<  \left( \begin{array}{ll}
         \delta_g^s(\mathbf{k},z)  \\
         \delta_{GW}(\mathbf{k},z)
    \end{array}\right)
    \left( \delta_g^s(\mathbf{k'},z)\: \delta_{GW}(\mathbf{k'},z)  \right)
    \right>  = (2\pi)^3 \delta_D (\mathbf{k} -\mathbf{k'})\left( 
    \begin{array}{cc}
P_{gg}(\mathbf{k},z) + \bar{n}_g(z)^{-1} & P_{g\; GW}(\mathbf{k}, z)    \\
P_{g\; GW} (\mathbf{k},z) &  P_{GW\; GW}(\mathbf{k},z) + \bar{n}_{GW}(z)^{-1}
\end{array}
    \right),\label{covmat}
    \end{align}
\end{widetext}
where $P_{xy}(\mathbf{k},z)$ is the three dimensional power spectrum at redshift $z$ associated with clustering between two traces $(\lbrace x,y \rbrace \in \lbrace g,GW \rbrace)$, $\delta_D(\mathbf{k}-\mathbf{k'})$ is the Dirac delta function and $\bar{n}_x(z)^{-1}$ represents the shot noise contribution, non-zero only for $x=y$. We can express  $P_{xy}^{ij}(\mathbf{k},z)$ in terms of the matter power spectrum $P_m(k,z)$, \citep{Mukherjee:2020hyn},  as
\begin{align}
    \nonumber  P_{gg}(\mathbf{k},z) &= b_g^2(k,z) (1+ \beta_g \mu^2_{\hat{k}})^2P_m(k,z), \\[10pt]
   \nonumber  P_{g\; GW}(\mathbf{k},z) &= b_g(k,z)b_{GW}(k,z)(1+ \beta \mu^2_{\hat{k}}) P_m(k,z), \\[10pt]
    P_{GW\; GW}(\mathbf{k},z) &= b^2_{GW}(k,z)P_m(k,z).
    \label{power spectrum}
\end{align}

{The spatial distribution of the galaxies will follow the underlying dark matter distribution. As a result, it will be related to the dark matter power spectrum by the bias parameter. The spatial correlation between the GW sources and galaxies arises because the GW sources will form in galaxies. So, it will be a different biased tracer of the same underlying dark matter distribution. The non-zero value of this cross-correlation signal will help in identifying the clustering redshift of the GW sources.} We will use this technique to explore the distance-redshift relationship to measure the cosmological parameters from the dark GW sources.  {In Eq. \eqref{covmat} we show the joint power spectrum between the galaxy-galaxy, GW-galaxy, and GW-GW part. The cosmological information in the GW-GW part is sub-dominant than the shot noise component. Whereas the maximum statistical information is there in the galaxy-galaxy side due to observation of a large number of galaxies. The cross-correlation power spectrum between a large number of galaxies and GW sources helps in extracting the cosmological redshift for the GW source. In the remaining of the paper, we will only focus on the cross-correlation between the GW sources and galaxies and the auto power spectrum part of galaxies.}  


\section{Important factors to explore the synergies between galaxy surveys and dark standard sirens}\label{synergy}
 We will mention the key aspects which play a crucial role in the implementation of the cross-correlation technique. 

\begin{figure*}
    \centering
    \includegraphics[width=1.\linewidth]{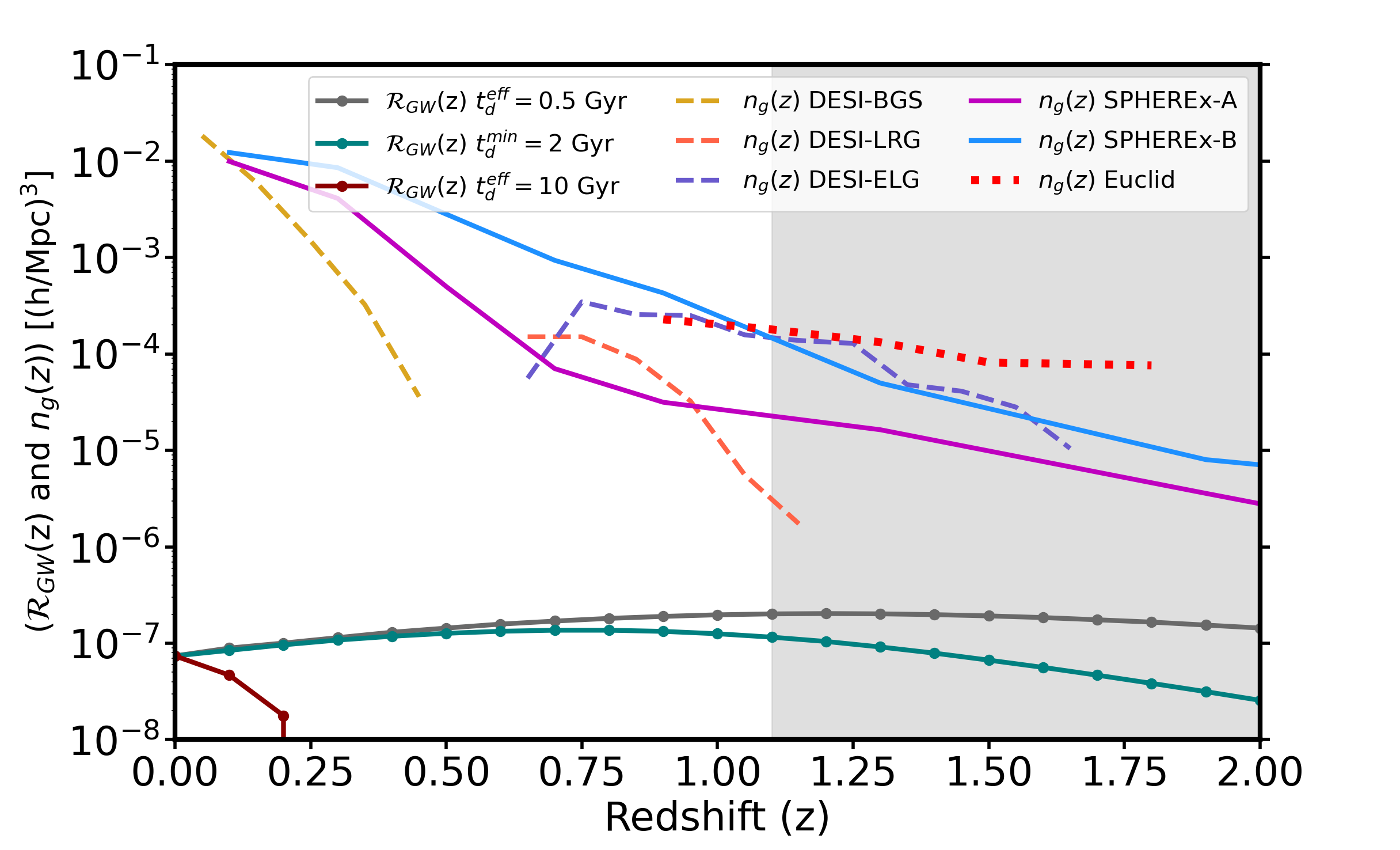}
    \captionsetup{singlelinecheck=on,justification=raggedright}
    \caption{We show the overlap in redshift between the different galaxy surveys along with the possible GW mergers for three different delay time model integrated over one year observation time. The shaded region in grey denotes the redshift for which detection beyond network matched-filtering signal to noise ratio greater than eight is not possible for most of the sources.}\label{fig:overlap}
\end{figure*}

\textit{Sky localization error:} The accuracy and precision of the cross-correlation technique depend on the amount of small-scale (or large values of Fourier modes $k$) clustering information that can be utilized. One of the parameters which control the maximum value of Fourier modes $k$ that can be used in the analysis is the sky localization error. The larger angular size of the sky localization error $\Omega_{GW}$ leads to a smoothing of the cross-correlation power spectrum for comoving Fourier modes larger than $k^p_{max}(z)\equiv \frac{\sqrt{8\ln 2}}{\sqrt{\Omega_{GW}(z)}d_c(z)}$ perpendicular to the line of sight, where $d_c(z)$ is the comoving distance at a redshift $z$ \citep{Mukherjee:2020hyn}. As a result, if the value of $\sqrt{\Omega_{GW}}d_c$ is large, then $k^p_{max}$ is small which will lead to a suppression in the cross-correlation estimation for values of Fourier modes $k$ larger than $k^p_{max}$. As a result, GW sources that are nearby and have better sky localization can provide a more precise measurement of the clustering redshift than the GW sources which are farther out. So to achieve a better sky localization, we need a network of GW detectors \citep{KAGRA:2013pob, Grover:2013sha, Howell:2017wvf, Chan:2018fpv} with high detector sensitivity. 

\textit{Luminosity distance uncertainty :} The uncertainty on the luminosity distance to the GW sources plays a crucial role in the cross-correlation technique. The presence of luminosity distance uncertainty causes smearing of the correlation function along the line of sight. For the Fourier modes larger than $k^l_{max}\equiv (1+z)/\sigma_{d_l}$, the contribution to the contribution to the cross-correlation signal gets suppressed for comoving Fourier modes greater than $k^l_{max}$. As a result, sources with smaller luminosity distance errors are more useful for measuring the cosmological parameters using the cross-correlation technique than the sources having larger luminosity distance errors.

\textit{Overlapping sky area between GW sources and galaxy surveys:}
The implementation of the cross-correlation between the GW sources and galaxies requires to have overlapping sky footprints. A network of GW detectors can detect sources from all the sky directions, though with different sensitivities in different directions depending on the position of the detectors. However, for the galaxy surveys, full-sky coverage from a single mission is possible only from space-based detectors, and not from the ground. From the ground (depending on the location of the detector), we can map only a fraction of the sky. However, by combining the data from multiple telescopes, we can enhance the total footprint of the galaxy surveys. So, implementation of the cross-correlation technique will be best for a full sky-mission or combinations of several missions to cover a large sky fraction.

\textit{Accurate redshift estimation of galaxies:}
The estimate of the clustering redshift of the GW sources by the cross-correlation technique requires an accurate and precise measurement of the galaxy redshift. So, synergies with GW sources can be fruitful with galaxies having spectroscopic redshift measurements. The spectroscopic measurement of the galaxy redshift can give us an unbiased redshift estimation and also uncertainty $\sigma_z/(1+z)$ around $\sim 10^{-3}$.  For galaxy surveys with photometric redshift measurements, the uncertainties on redshift can be of the order of a few percent ($\geq 2\%$) and may not be completely accurate. So, in this analysis, we will only explore the synergies with the upcoming/ongoing spectroscopic galaxy surveys. The possibility of using photometric galaxy surveys will be explored in future work. 

\textit{Redshift distribution of galaxies:} The cross-correlation of the GW sources with galaxy relies on the availability of galaxies samples up to high redshift. {One of the key advantages of the cross-correlation technique is not dependent on whether the host galaxy of the GW sources is present in the galaxy catalog or not \citep{Mukherjee:2018ebj, Mukherjee:2020hyn, Bera:2020jhx}}  {In the statistical host identification technique \citep{PhysRevD.101.122001}, the method depends on whether the host galaxy of GW source is present in the galaxy catalog}. The cross-correlation study depends on the number density of the galaxy as a function of redshift $n_g(z)$.  {For large values of $n_g(z)$, the shot noise contribution reduces. As a result, the total signal-to-noise ratio (SNR) of the cross-correlation signal increases. So, irrespective of whether the host galaxy is present or not in a galaxy catalog, clustering redshift can be inferred.} This is useful for an accurate estimate of the clustering redshift and also reducing the error on the redshift estimation. So, galaxy surveys with higher galaxy number density are essential to explore cross-correlation with GW sources. 

\section{Ongoing/Upcoming surveys of GW and galaxies  which can make the cross-correlation study feasible}\label{surveys}
We discuss the ongoing/upcoming detectors in the field of GW and galaxy surveys that can play a crucial role in inferring the cosmological parameters. We show in Fig. \ref{fig:overlap} the overlap in the redshift for different spectroscopic galaxy surveys with the Gw sources with different delay times. The shaded region in Fig. \ref{fig:overlap} denotes the redshift range that is inaccessible from the network of \texttt{LVK} detectors.  

\subsection{Network of GW detectors}\label{lvk}
A network of GW detectors helps in improving the sky localization error and reduces the uncertainty in the luminosity distance. As a result, the cross-correlation study becomes effective with a network of GW detectors. We consider a network of four GW detectors namely the two \texttt{LIGO} detectors (\texttt{LIGO-Hanford} and  \texttt{LIGO-Livingston}) \citep{aLIGO}, \texttt{Virgo} \citep{Acernese_2014}, and \texttt{KAGRA} \citep{KAGRA:2018plz} in their design sensitivity. Currently, \texttt{LIGO-Hanford}, \texttt{LIGO-Livingston}, and \texttt{Virgo} are operational, and from 2022, the \texttt{KAGRA} will likely be operational. We have used five years of observation time with a $50\%$ duty cycle in our analysis. Further improvement in the sky localization error is possible with the operation of a fifth GW detector, \texttt{LIGO-India} \citep{Unnikrishnan:2013qwa}. We will not consider \texttt{LIGO-India} in this work and will defer the study to a future study. 

\subsection{Spectroscopic galaxy surveys}\label{emprobes}
To be able to perform the cross correlation technique, we need galaxy surveys which satisfy the criteria mentioned in Sec. \ref{synergy}. One of the crucial requirements for the cross-correlation method is to have accurate redshift estimation to galaxies. So, we have explored the possible synergy with the spectroscopic surveys such as \texttt{DESI} \citep{desi}, \texttt{SPHEREx} \citep{spherex}, and \texttt{EUCLID} \citep{laureijs2011euclid} in this paper. We do not include \texttt{Roman Telescope} \citep{2012arXiv1208.4012G, 2013arXiv1305.5425S, Dore:2018smn} in this analysis, as the final design is still at the stage of preparation.  In a future work, we will be exploring the synergy between \texttt{Roman Telescope}  \citep{2012arXiv1208.4012G, 2013arXiv1305.5425S, Dore:2018smn} and \texttt{Vera Rubin Observatory} \citep{2009arXiv0912.0201L} with the \texttt{LVK} detectors.  

\subsubsection{\texttt{DESI}}
\texttt{DESI} \citep{desi} is a ground-based dark-energy experiment designed to study baryon acoustic oscillations (BAO) and the growth structure through redshift-space distortions with a wide-area galaxy and quasar redshift survey. It will conduct a 5-year survey designed to cover 14,000 deg$^2$, which corresponds to a sky fraction $f_{sky}= 0.3$. \texttt{DESI} will survey at $0.4 < z < 3.5$ using luminous red galaxies (LRGs), emission-line galaxies (ELGs), and quasars, to measure the large-scale clustering of the Universe: they will measure  LRGs up to $z=1$, ELGs up to $z = 1.7$ and quasar up to $z=3.5$. It will also carry out a Bright Galaxy Survey (BGS) of the $z<0.4$ Universe, which will consist of approximately 10 million galaxies. In total, they will make more than 30 million galaxies and quasars. \texttt{DESI} will use a combination of three telescopes to provide the baseline targeting data. The first of their targets is the Bright Galaxy Sample (BGS) that will have a redshift distribution peak around $z = 0.2$. As for LRGs, the expected redshift peaks at redshift $z< 0.8$.  The ELGs redshift distribution peaks at around $z=0.8$ and it does not drop to a level where shot noise dominates BAO measurements until $z \approx 1.3$. Lastly, the expected redshift distribution for QSO peaks at around $z \approx 1.6$.

In our work, we are covering the redshift range from $z=0.1$ to $z=0.5$. Taking into account the galaxy density distributions in Fig.  \ref{fig:overlap}, we can see that there will be an overlap in redshift between our analysis and \texttt{DESI}'s LRGs measurements in the higher redshift range of our analysis and with BGS measurements up to around $z=0.4$.

\subsubsection{\texttt{SPHEREx}}
\texttt{SPHEREx} \citep{spherex, Dore:2018kgp} is an all-sky spectroscopy survey satellite designed to probe the origin and density of the Universe. It focuses on the low-redshift Universe and aims to determine the redshifts of hundreds of millions of galaxies. The brightness of the galaxy strongly correlates with the redshift error. SPHEREx expects to measure approximately 9 million galaxies with 0.3$\%$ accuracy up to $z \approx 0.6$. We can see in Fig.  \ref{fig:overlap} the observed galaxy density for the redshift error $\sigma_z=$ 0.003$(1+z)$ and $\sigma_z=$ 0.01$(1+z)$ respectively shown in magenta (\texttt{SPHEREx-A}) and blue (\texttt{SPHEREx-B}). We restrict our analysis to these redshift bins because high-accuracy redshift measurements are necessary to allow for clustering measurement and thus apply the cross-correlation technique.

\texttt{SPHEREx} derives the photometric redshift of galaxies extracting the flux from simulated images based of galaxy spectral density energy distribution (SEDs) from real data and then fitting template SEDs to the observations. This gives the redshift probability distribution function (PDF), $P(z)$, from which they obtain the best redshift estimate $\bar{z} = \frac{\sum_{i} z_iP(z_i)}{\sum_{i}P(z_i)}$ and the error on the estimate $\sigma_{\bar{z}} =\sqrt{\frac{\sum_{i} (z_i - \bar{z})^2 P(z_i)}{\sum_{i} P(z_i)}}$ \citep{spherex}.

\subsubsection{\texttt{Euclid}}
\texttt{Euclid} \citep{laureijs2011euclid,Euclid:2019clj} will cover 15,000 deg$^2$ of the northern and southern extragalactic sky and will measure spectroscopic redshifts of 30 million galaxies galaxies in the redshift range $0.9 < z < 1.8$ with an spectroscopic redshift error of $0.001(1+z)$. The expected number density of galaxies, i.e of observed H$\alpha$ emitters, for each redshift bin in the given redshift range is of the order of $10^{-4}$ ($h^3$Mpc$^{-3}$), as we can see in Fig.  \ref{fig:overlap}. So, the overlap between the \texttt{Euclid} galaxy samples and the GW sources will be only at high redshift $0.9<z<1.1$. For these sources, the luminosity distance error and sky localization error will be large for \texttt{LVK} network of detectors. As a result, the cross-correlation technique will not be effective. However, it is important to include \texttt{Euclid} in the combined catalog of galaxy surveys, to allow cross-correlation over a large redshift range. In future, with the operation of the next generation GW detectors such as Laser Interferometer Space Antenna (\texttt{LISA}) \citep{2017arXiv170200786A}, \texttt{Einstein Telescope} \citep{Punturo:2010zz}, \texttt{Cosmic Explorer} \citep{Reitze:2019iox} and \texttt{TianQin observatory} \citep{Luo:2015ght}, cross-correlation with \texttt{Euclid} galaxy survey will be very useful.  

\begin{figure}
    \centering
    \includegraphics[width=1.\linewidth]{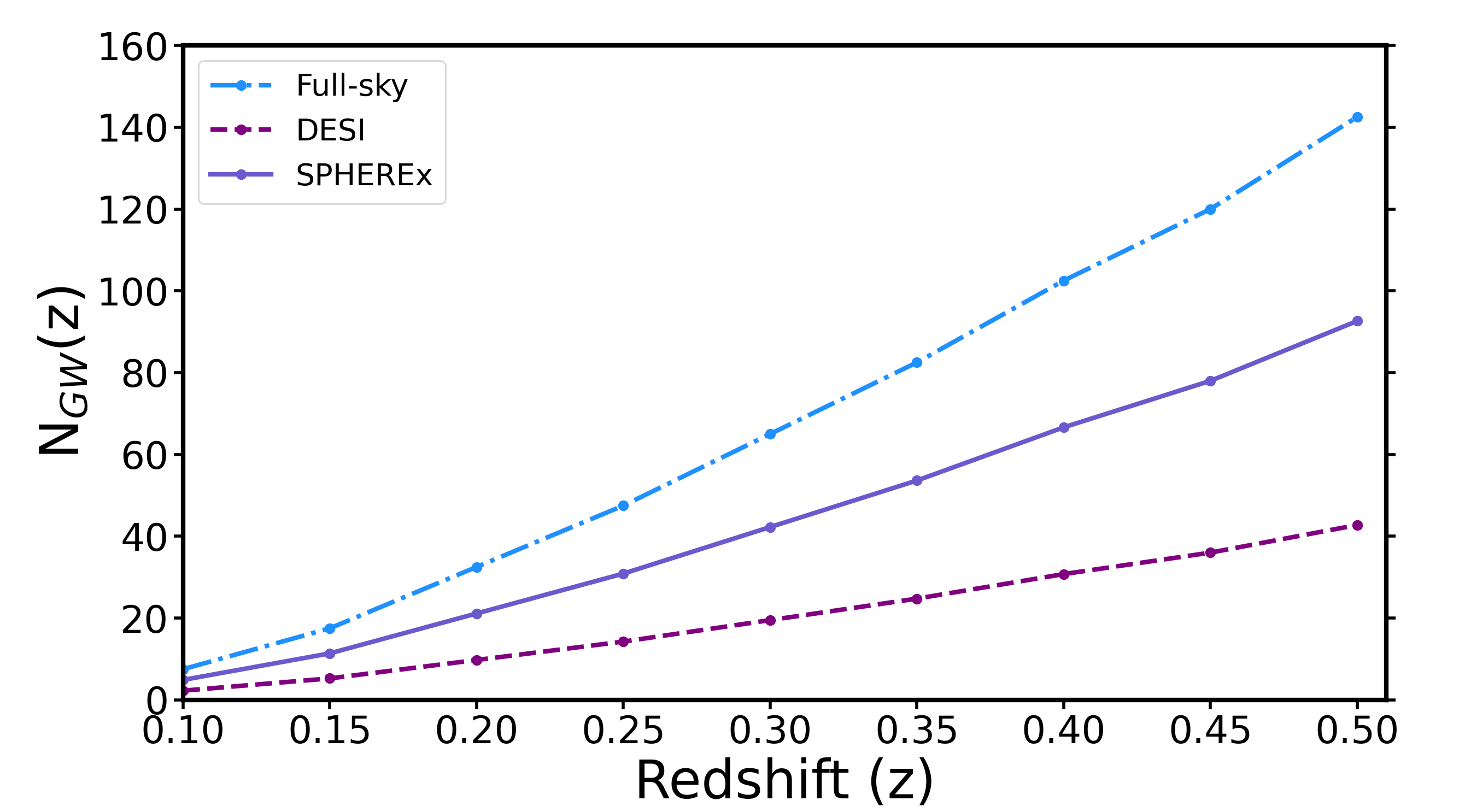}
    \caption{Number of GW sources in  full-sky, in the \texttt{DESI} sky area, and in the \texttt{SPHEREx} sky area as a function of redshift $z$ for the power-law delay time parameter $t_d^{min} = 0.5$ Gyr , for 5 years of observation time with 50$\%$ duty cycle.}
    \label{fig:modelC_sources}
\end{figure}

\section{ {Bayesian framework for the inference of the cosmological parameters}}\label{bayes}
\begin{figure*}
    \centering
    \includegraphics[width=1.\linewidth]{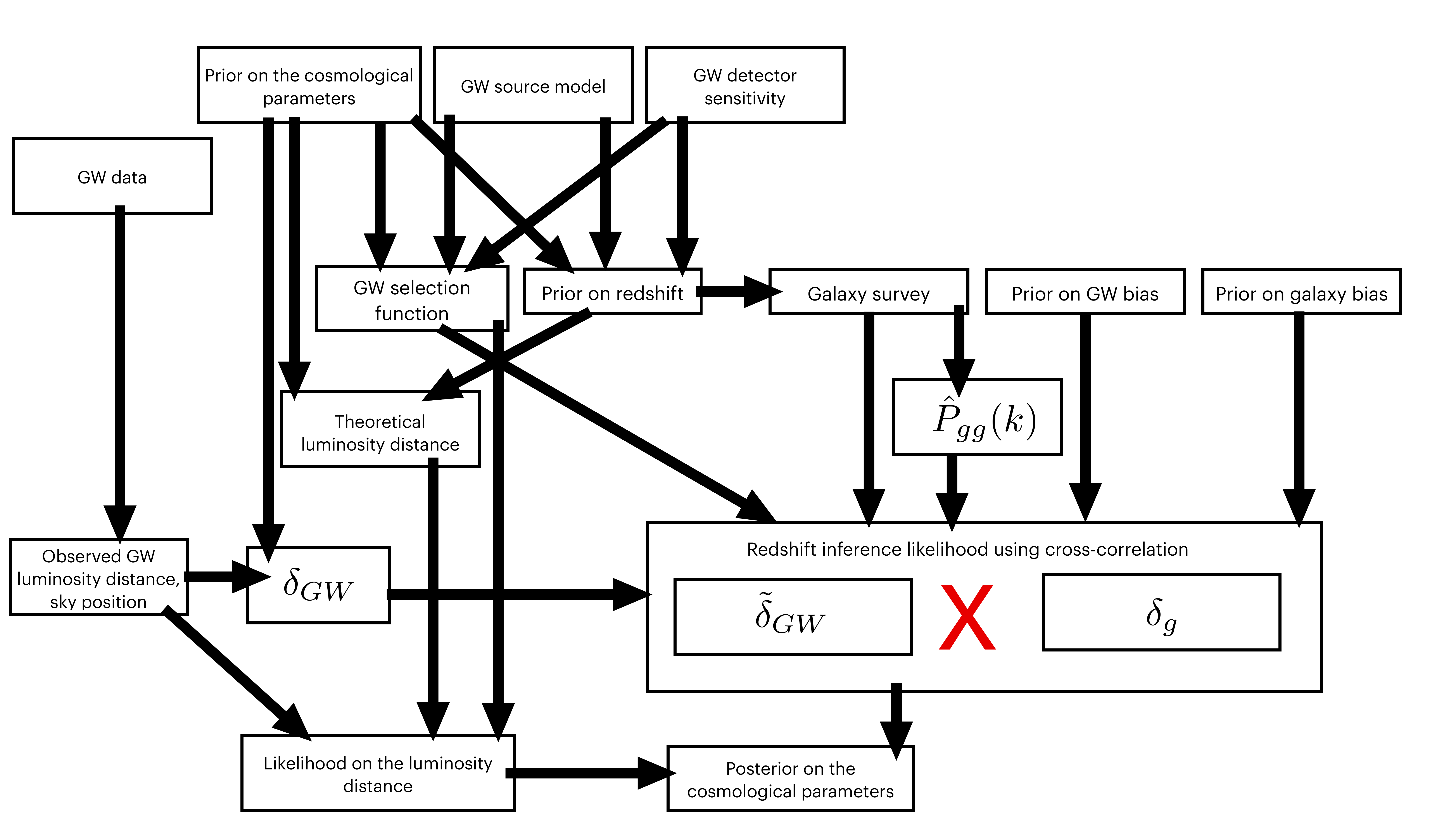}
    \captionsetup{singlelinecheck=on,justification=raggedright}
    \caption{ {We show a network diagram of the Bayesian framework for the implementation of cross-correlation between GW sources and galaxies to infer the cosmological parameters by exploring the luminosity distance and redshift relation using the dark standard sirens.}}\label{fig:flowchart}
\end{figure*}

We discuss here the framework to evaluate the cosmological parameters from dark standard sirens using the cross-correlation technique. This technique relies on the estimation of the clustering redshift of the GW sources using cross-correlation with galaxies having spectroscopic/photometric redshift measurements. The GW sources exhibit more spatial clustering with the galaxies at the correct redshift than with the other galaxies. As a result by using this technique, we can infer the clustering redshift and in combination with the luminosity distance estimation of the GW sources, we can measure the cosmological parameters. As the GW sources are much less in number than the galaxies (see Fig. \ref{fig:overlap}), the main statistical power comes from a large number of galaxies. As a result, the statistical power in the GW-GW correlation part is negligible. So we primarily use the cross-correlation part between the GW-galaxies and the auto power spectrum between the galaxies in this framework. A network diagram explaining this framework is shown in Fig. \ref{fig:flowchart}. I will describe this framework in detail.  

The posterior on the cosmological parameters denoted by $\Theta_c$ can be written using the Bayes theorem for the given GW data $\vec{\vartheta}_{GW}$, the galaxy samples $\vec d_{g}$ as
\begin{equation}\label{eqbayes1}
    \mathcal{P}(\Theta_c|\vec{\vartheta}_{GW}, \vec d_{g})= \frac{P(\vec{\vartheta}_{GW}|\vec d_{g}, \Theta_c)\Pi(\Theta_c)}{\mathcal{E}}
\end{equation}
where $P(\vec{\vartheta}_{GW}|\vec d_{g},\Theta_c)$ is the likelihood,  $\Pi(\Theta_c)$ is the prior on the cosmological parameters and $\mathcal{E}$ is the evidence. The GW data vector is composed of the posterior of the luminosity distance and sky position inferred from the GW data. The GW sources are distributed over nearly the whole sky and up to a maximum luminosity distance which is given by the luminosity distance posterior obtained from the GW data. So, we can construct a space of the luminosity distance from zero to the maximum value of the luminosity distance $d_l^{\rm max}$ allowed by the GW network of detectors for an allowed range of source parameters such as component mass, spin, inclination angle, sky position, etc. for a particular choice of the population model. With a given choice of the cosmological parameters prior, the maximum value of the luminosity distance $d_l^{\rm max}$ translates to a maximum redshift which can be denoted by $z_{\rm max}$. So, the allowed range of prior on $\Pi(z)$ is from zero to $z_{\rm max}$. So, $z_{\rm max}$ is decided by the best possible sources that are detectable up to the highest redshift for the extreme value allowed by the prior on the cosmological parameters. In principle, the value of $z_{\rm max}$ can be taken arbitrarily large, but then the detectability of the sources beyond $z_{\rm max}$ will be zero.  Now we would like to write the Bayesian framework to implement the cross-correlation between the GW sources and galaxies. The selection function of GW sources above a certain matched filtering SNR is not written explicitly in the likelihood as it does not contain anything related to the cross-correlation technique, so we absorb it in the proportionality constant in Eq. \eqref{eqbayes2}. The selection function decides all the possible GW sources that are detectable above a matched filtering SNR threshold $\rho_*$ for a given set of cosmological parameters.  Considering galaxy samples with spectroscopic (or photometric) redshift measurement we can factorize the likelihood of Eq. \eqref{eqbayes1} as

\begin{align}\label{eqbayes2}
    P(\vec{\vartheta}_{GW}, \vec d_{g}|\Theta_c)&\propto \prod^{N_{GW}}_i\int dz  \iint d\theta d\phi \mathcal{S}(\theta, \phi) \nonumber \\ & \times P(d^i_l|z, \Theta_c,\{\theta^i, \phi^i\}_{GW}) \nonumber \\ & \times P(\delta_{GW}(\hat z(d_l,\Theta_c),\{\theta^i, \phi^i\}_{GW})|\vec d_{g})\Pi(z), 
\end{align}
where $N_{\rm GW}$ denotes the number of GW sources which are detected above a certain matched filtering SNR. The terms in the above equation such as $\mathcal{S}(\theta, \phi)$ denotes the sky localization error of the GW sources,  $P(d^i_l|z, \Theta_c,\{\theta^i, \phi^i\}_{GW})$ denotes the probability of getting observed luminosity distance given the redshift and the cosmological parameters, $P(\delta_{GW}(\hat z(d_l,\Theta_c),\{\theta^i, \phi^i\}_{GW})|\vec d_{g})$ denotes the probability of getting the GW source given the galaxy distribution $\vec d_{g}$  which is written in terms of the GW density field $\delta_{GW}(\hat z(d_l,\Theta_c),\{\theta, \phi\}_{GW}$ that is obtained by transforming from the observed luminosity distance to the redshift space assuming a set of cosmological parameters from the prior $\Pi(\Theta_c)$, and $\Pi(z)$ denotes the prior on the allowed redshift range. Since the sky localization error for the GW sources are very large, we need to convolve the GW density field with the sky map of the GW sources (denoted by $\mathcal{S}(\theta, \phi)$ which is included in terms of the integration over the sky area $\theta$ and $\phi$.
Eq. \eqref{eqbayes2} can now be in written after integrating over the sky localization error as 
\begin{align}\label{eqbayes2a}
    P(\vec{\vartheta}_{GW}, \vec d_{g}|\Theta_c)& 
    \propto \prod^{N_{GW}}_i\int dz  P(d_l|z, \Theta_c,\{\hat \theta^i, \hat \phi^i\}_{GW})\nonumber \\ & \times  P(\tilde \delta_{GW}(\hat z(d_l, \Theta_c),\{\hat \theta^i, \hat \phi^i\}_{GW})|\vec d_{g})\Pi(z),
\end{align}
where we write the GW density field in terms of the convolved density field $\tilde \delta_{GW}(\hat z(d_l,\Theta_c),\{\hat \theta^i, \hat \phi^i\})$ after integrating over the sky localization error. The term $P(d_l|z, \Theta_c,\{\hat \theta^i, \hat \phi^i\}_{GW})$ is now written for the luminosity distance after marginalizing over the sky position. The quantity $P(\tilde \delta_{GW}(\hat z(d_l,\Theta_c),\{\hat \theta^i, \hat \phi^i\}_{GW})|\vec d_{g})$ captures the probability of getting GW density field given the galaxy density field. This quantity explores the spatial clustering of the GW sources with galaxies. 

For every choice of the cosmological parameters, we construct a GW density field in the redshift space, and then cross-correlate that with a galaxy distribution distributed up to the $z_{\rm max}$ value allowed by the prior on redshift $\Pi(z)$. The cross-correlation is implemented over multiple redshift bin widths $\Delta z$ between the GW density field and the galaxy density field. The exact choices of the bins depend on the properties of galaxy surveys such as their redshift errors. We construct the cross-correlation signal in Fourier-space with every Fourier mode denoted by $k$. \footnote{The cross-correlation can also be implemented in the real space or the spherical harmonic space.} We write the above equation for different Fourier modes. As the individual Fourier modes are independent, we can combine the power spectrum estimation for all different Fourier modes $k$ by taking the product of all independent Fourier modes as
\begin{align}\label{eqbayes3}
    P(\vec{\vartheta}_{GW}, \vec d_{g}|\Theta_c)& \propto \prod^{N_{GW}}_i\int dz  P(d_l|z, \Theta_c,\{\hat \theta, \hat \phi\}_{GW}) \int d\Theta_n \nonumber\\ & \times \prod_k P(\hat P_{gGW}(k,z)| P_{m}(k,z), b_g(z), b_{GW}(z))\nonumber \\ & \times P(\delta_g, z| P_{m}(k,z), b_g(z))\Pi(\Theta_n)\Pi(z),
\end{align}
where, $\Theta_n \in \{b_g(z), b_{GW}(z)\}$ denotes the nuisance parameters, $\hat P_{gGW}(k,z)$ is the measured power spectrum between GW density field and galaxy density field, and $P_{m}(k,z)$ denotes the matter power spectrum. Since, GW sources are going to trace the same realization of the dark matter density field (denoted by $\hat P_m(k,z)$) which will also be traced by the  galaxies, we expect the same realization matter power spectrum to be related to both galaxies and GW sources through the corresponding bias parameters i.e. $\hat P_{gGW}(k,z)= b_{g}(z)b_{\rm GW}(z)\hat P_m(k,z) = \frac{b_{\rm GW}(k,z)}{b_g(k,z)}\hat P_{gg}(k,z)$. So, we use the constrained realization of the matter power spectrum which is related to the auto-correlation of the galaxy density field in the analysis. This is denoted by the third term in Eq. \eqref{eqbayes3}  $P(\delta_g, z| P_{m}(k,z), b_g(z))$. This part takes into account the information coming from the first block of the matrix given in Eq. \eqref{covmat}. 

Since the galaxy and GW bias parameters are unknown, we need to marginalize them in the analysis. The bias parameters can also be scale-dependent in the non-linear scales.  But as the sky localization error of the GW sources is very large, we do not resolve the non-linear scales where the bias parameter can be scale-dependent. As a result, one can assume a scale-independent bias parameter in the linear scales. This is discussed in detail in the previous section. The GW bias parameter depends on the types of sources detected by the GW detectors and also on the selection function of the GW sources. GW sources of certain masses are only detectable at high redshift. As a result, the corresponding bias parameter going to depend on the properties of sources such as their component masses, inclination, spin, etc. As the auto-correlation of the galaxy density field captures the galaxy power spectrum, we can write the above equation as 
\begin{align}\label{eqbayes4}
    P(\vec{\vartheta}_{GW}, \vec d_{g}|\Theta_c)& \propto \prod^{N_{GW}}_i\int dz  P(d_l|z, \Theta_c,\{\hat \theta, \hat \phi\}_{GW}) \int d\Theta_n \nonumber\\ & \times \prod_k P(\hat P_{gGW}(k,z)| \hat P_{m}(k,z), b_g(z), b_{GW}(z))\nonumber \\ & \times P(\delta_g, z| \hat P_{gg}(k,z))\Pi(\Theta_n)\Pi(z).
\end{align}
Now, we assume that at the linear scales, we can write the likelihood $P(\hat P_{gGW}(k,z)| P_{m}(k,z), b_g(z), b_{GW}(z))$ for each Fourier modes as a Gaussian with variance $C(k,z)$. Then the product of different Gaussian terms can be written as the exponential of the integration over different Fourier modes which we denote in terms of the cross-correlation likelihood $\mathcal{L}$ as
\begin{align}\label{eqbayes5}
    P(\vec{\vartheta}_{GW}, \vec d_{g}|\Theta_c)& \propto \prod^{N_{GW}}_i\int dz  P(d_l|z, \Theta_c,\{\hat \theta, \hat \phi\}_{GW})\nonumber \\ & \times \int \Theta_n \mathcal{L}(\vec{\vartheta}_{GW}| \{P_{gg}(z)\}, \Theta_n, \vec d_g(z))\nonumber \\ & \times P(\delta_g, z| \{P_{gg}(z)\}, \Theta_n)\Pi(\Theta_n)\Pi(z).
\end{align}
The cross-correlation likelihood in Eq. \eqref{eqbayes5}   $\mathcal{L}(\vec{\vartheta}_{GW}| \{P_{gg}(z)\}, \Theta_n, \vec d_g(z))$ can be written in terms of the overlapping sky volume $V_s$ between the GW sources and galaxies,  the cosine of the angle between the line of sight and the Fourier mode $\hat{k}$ $\mu$ as \citep{Mukherjee:2020hyn}\footnote{The notation $\{P_{gg}(z)\}$ in the term $\mathcal{L}(\vec{\vartheta}_{GW}| \{P_{gg}(z)\}$ is only to denote a vector that includes the power spectrum for different values of the Fourier modes $k$ at a redshift $z$.}
\begin{align}\label{likeli-1}
\begin{split}
  \mathcal{L}(\vec{\vartheta}_{GW}| &\{P_{gg}(z)\}, \Theta_n, \vec d_g(z))  \propto  \exp\bigg(-\frac{V_s}{4\pi^2}\int k^2 dk \int d\mu\\ & \times {\mathbf{D}^\dagger(k,\mu, z)\mathbf{C}^{-1}(k, \mu,z)\mathbf{D}(k, \mu,z)}\bigg),
  \end{split}
\end{align}
where $\mathbf{D}(k,\mu,z)\equiv \hat P (\vec k, \Delta \Omega_{GW}) - b_g(k,z)b_{GW}(k, z)(1 + \beta_g \mu_{\hat k}^2)\hat P_{m}(k,z)e^{-\frac{k^2}{k^2_{\rm eff}}}$, consists of the power spectrum estimated from data ($\hat P (\vec k, \Delta \Omega_{GW})$) and the model power spectrum ($b_g(k,z)b_{GW}(k, z)(1 + \beta_g \mu_{\hat k}^2)\hat P_{m}(k,z)e^{-\frac{k^2}{k^2_{\rm eff}}}$). The value of effective Fourier modes up to which signal can be estimated is $1/k^2_{eff}= \mu^2/(k^{l}_{max})^2 + (1-\mu^2)/(k^p_{max})^2$, where the value of $k^{p}_{max}$ and $k^{l}_{max}$ are defined in Sec. \ref{synergy}.  {The value of $k^p_{max}$ depends on the measured quantities such as the error on the luminosity distance and sky localization uncertainty and on the redshift bin with which cross-correlation is performed. It does not depend on the choice of the cosmological parameters. As a result, the value of $k^p_{max}$ does not impact the mean value of the inferred cosmological parameters. It only affects the error-bar of the inferred cosmological parameters and the bias parameters.}  {The Gaussian form of the smoothing kernel is valid only when the sky localization error is Gaussian (or close to Gaussian). For the network of four LVK detectors, the sky localization error posteriors can be well approximated as Gaussian for sources with a detector network SNR greater than 10. So, this approximation is well justified. However, if the GW localization error is significantly non-Gaussian, then one needs to convolve the non-Gaussian sky map with the power spectrum to estimate the corresponding model of the power spectrum. Though such an expression will not have a simple analytical form, it can be included in Eq. \eqref{likeli-1} straightforwardly.} 
The covariance matrix considered in this analysis is $\mathbf{C}(k,\mu,z)\equiv2(P_{gg}(\vec k,z) + n_g(z)^{-1})(P_{GW\,GW}(\vec k,z) + n_{GW}(z)^{-1})$ which is taken as diagonal in this analysis. At the large values of the Fourier modes $k$, the covariance matrix is also going to have off-diagonal terms. However, most of those scales are not resolvable due to the sky localization error of $10$ sq. deg or larger on the GW sources detectable from the \texttt{LVK} data analysis. So in this analysis, we ignore the off-diagonal terms.  {The form of the likelihood is considered to be Gaussian. It is commonly used for the analysis of several galaxy data sets \citep{10.1093/mnras/stv961, 2017MNRAS.470.2617A, Euclid:2019clj, 2021MNRAS.505..377R}.} 
However, for the implementation of the technique on the data, we can consider the full survey-specific covariance matrix.

\section{Simulated samples of GW sources and galaxies}\label{mock}
In this analysis, we do a Bayesian estimation of the GW source parameters using a nested sampler \texttt{dynesty} \citep{2020MNRAS.493.3132S} which is a part of the \texttt{Bilby} package \citep{bilby} for a network of four detectors \texttt{LIGO-Hanford}, \texttt{LIGO-Livingston}, \texttt{Virgo}, and \texttt{KAGRA}. This is a computationally time-consuming process and can be performed only for a few hundred sources \footnote{A single GW source parameter estimation (with six parameters) has taken $<10$ hours to about $50$ hours for heavier and lighter masses respectively.}. So, we will restrict our analysis only to the low redshift ($z\leq 0.5$) to explore the cross-correlation study. This choice gives us a limited number of samples, for which we perform GW parameter estimation and propagate the uncertainties properly in the cross-correlation study to gauge its utility. As a result, we will be able to explore the synergies with only the mock catalog for low redshift BGS samples from a \texttt{DESI} survey and with the low-redshift galaxies from a \texttt{SPHEREx} survey. However, this choice is unlikely to make any major impact on the analysis. Because the sources at higher redshift are going to have low matched-filter SNR than the nearby samples that will cause a larger luminosity distance error and also smaller a smaller value of $k_{eff}$. As a result, it will make the estimation of the clustering redshift weak. The only positive aspect of going to high redshift is an increase in the total number of sources, which can improve the measurement of cosmological parameters. So, in a future analysis, we will use approximate-likelihood to consider GW sources up to high redshift to explore the synergy with \texttt{DESI-ELG}, \texttt{Vera Rubin Observatory}, and \texttt{Roman Telescope}. In this analysis, we will also not consider the spin parameters to reduce the computational time of the parameter estimation. However, this assumption is also not going to make any major impact on the cross-correlation study. 

\subsection{GW source catalog} \label{GWcatalog}
We consider six parameters for preparing the GW catalog. These parameters are the component masses of the two black holes $m_1$ and $m_2$, the luminosity distance to the GW source $d_l$, the inclination angle $\theta_{\text{jn}}= \arccos(\hat L.\hat n)$ (the angle between the angular momentum vector $\hat L$ and the line of sight $\hat n$), and the sky location Right ascension (RA) and declination (DEC). Also, to generate a GW mock catalog, we need to consider a redshift distribution of the GW merger rates.  We will briefly describe below all these aspects which we use to construct a mock GW catalog. 

\textit{GW Merger rate:} For making a mock catalog of GW sources, we need to assume a merger of GW sources as a function of redshift. We consider a GW merger rate model motivated by the \citet{Madau2014} SFR and with power-law delay time distribution with a minimum delay time, $t_d^{eff}= 0.5$ Gyr. The corresponding merger rate is shown in Fig. \ref{fig:plot_powerlaw} in grey line.  We consider a total observation time of five years with a $50\%$ duty cycle with a network of \texttt{LVK} detectors and perform a integral over the comoving volume with a redshift bin-width $\Delta z=0.02$ and observation time to obtain the total number of samples $N_{GW}(z)= \int dt \int dz \frac{\mathcal{R}_{GW}(z)}{1+z} \frac{dV}{dz}$. The corresponding plot for the total number of sources is shown in Fig. \ref{fig:modelC_sources} in a dashed-dotted line.

\textit{GW source mass distribution:} In this analysis we assume a power-law mass distribution of black holes. A power-law model for the BH mass distribution is used in LIGO data analysis, \citep{Abbott_2019, Abbott2016a}. More recently, a power-law+peak model was found to be more suitable to the data with a peak around a mass of $30-40$ M$_\odot$. The presence of such a peak provides additional cosmological information \citep{Farr:2019twy, Mastrogiovanni:2021wsd} and also for heavier mass black holes, we can expect a better measurement of the luminosity distance than a lighter one. However, in this analysis, we have taken a pessimistic choice on the mass distribution.  We consider a power-law distribution $M^{-\alpha}$ with a minimum mass $M_{min}=5$ M$_\odot$ and a maximum mass $M_{min}=50$ M$_\odot$ with a value of $\alpha=2.35$.  The lower mass limit in the mass is motivated by the lower mass gap. As for the upper limit, $m_{max} = 50M_\odot$ is determined by the lower edge of the pair-instability gap in the BBH mass spectrum \citep{Abbott2016a, Farmer}.  The peak in the mass distribution can help to improve the measurement of the cosmological parameters. To generate GW sources from the power-law distribution, we randomly generate the masses of the component black holes according to the power-law mass distribution using the acceptance-rejection method. This is performed for all the sources estimated $N_{GW}(z)$ for each redshift. 

\textit{GW inclination angle:} The inclination angle for the GW sources are considered randomly between $[0,\pi]$. For each redshift, we choose the number of $\theta_{JN}$ according to the value of $N_{GW}(z)$ obtained at each redshift. 

\textit{GW sky position:} The sky-area available for the cross-correlation study depends on the overlap between the GW sky-area (which can be considered full sky), and the galaxy surveys (which is survey specific). In this analysis, we consider two surveys for exploring the cross-correlation study, namely \texttt{DESI} and \texttt{SPHEREx}. The \texttt{DESI} BGS galaxy sample is available over a sky-area of 14000 sq. deg, which corresponds to a sky-fraction $f_{sky}=0.3$. For \texttt{SPHEREx}, in-principle the observation can be made over full-sky. But the effective sky-fraction $f_{sky}$ considered in this analysis is $0.65$.  {So, in our analysis, we randomly choose sky positions from a galaxy mock catalog. The number of GW sources depends on overlapping sky area between \texttt{LVK} with \texttt{DESI} and \texttt{SPHEREx}.} These are shown in Fig. \ref{fig:modelC_sources} in dashed-line and solid-line for \texttt{DESI} and \texttt{SPHEREx} respectively. 

\textit{Gravitational waveforms template:} For the generation of the GW signal, we use the \texttt{IMRPhenomPv2} \citep{IMRPhenom} gravitational waveform, which is available with the \texttt{Bilby} package \citep{bilby}. The \texttt{IMRPhenomPv2} \citep{IMRPhenom} waveform is generated in the frequency domain which is a  single-precessing-spin waveform model. This is model is composed of a post-Newtonian approximation to the inspiral period of the waveform and a phenomenological approximation for the merger and ring-down phase \citep{IMRPhenom}. 

\begin{figure*}
    \centering
    \includegraphics[width=1.\linewidth]{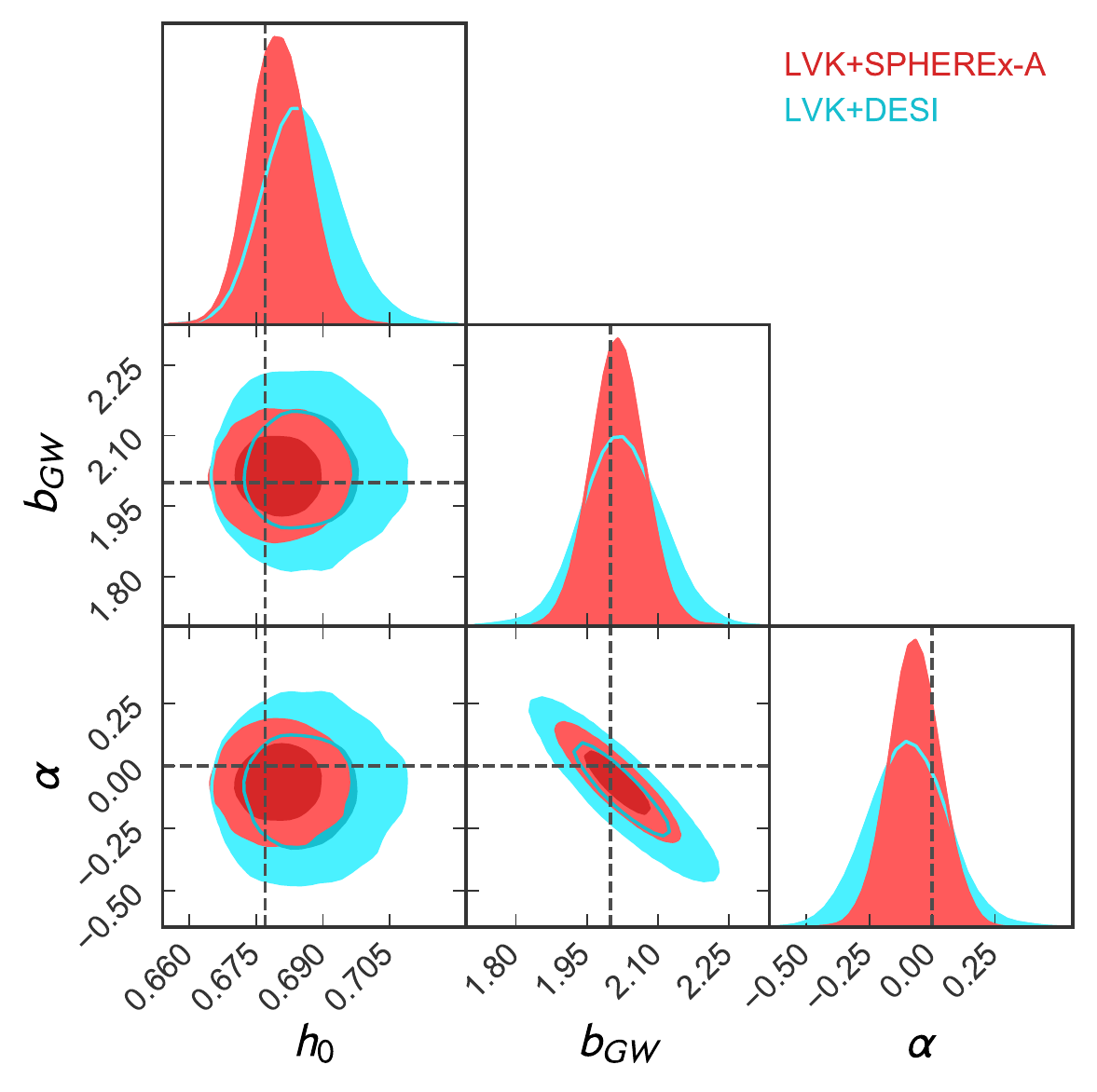}
    \captionsetup{singlelinecheck=on,justification=raggedright}
    \caption{Forecast for the joint estimation of the Hubble constant $H_0= 100 h_0$ km/s/Mpc along with the GW bias parameter $b_{GW}(z)= b_{GW}(1+z)^\alpha$ from the GW sources detectable from the network of \texttt{LVK} detectors in combination with \texttt{DESI} (cyan) and \texttt{SPHEREx-A} (red). The injected values in the simulations are shown by the black dashed line.}\label{fig:h0}
\end{figure*}

\begin{figure}
    \centering
    \includegraphics[width=0.9\linewidth]{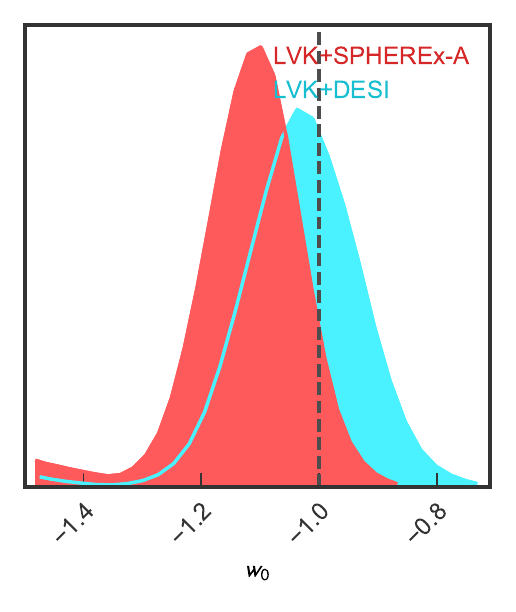}
    \captionsetup{singlelinecheck=on,justification=raggedright}
    \caption{The posterior on the dark energy equation of state $w_0$ after marginalizing over the GW bias parameter and its redshift dependence is shown for \texttt{LVK+DESI} (cyan) and \texttt{LVK+SPHEREx-A} (red). The black dashed-line indicates the injected value $w_0=-1$ used in the simulations.}\label{fig:w0}
\end{figure}

\subsection{Mock catalog of galaxy and GW sources}
For the study of the cross-correlation between dark standard sirens and upcoming galaxy surveys, we generate mock catalogs of galaxies with the redshift window function and number density of objects specific to individual large-scale structure surveys namely \texttt{DESI} \citep{desi}, and \texttt{SPHEREx-A} \citep{spherex, Dore:2018kgp}. The specifications for both these surveys are mentioned in Sec. \ref{emprobes}. For both these mock galaxy catalogs we use the package \texttt{nbodykit} \citep{Hand:2017pqn} including the effect of redshift space distortions. The maximum redshift is considered up to $z=0.7$ which is deeper than the GW source distribution considered in this analysis. The galaxy catalogs for \texttt{DESI} and \texttt{SPHEREx} are consider with spectroscopic redshift uncertainties. The galaxy samples are produced with a linear model of galaxy bias in this study with $b_g=1.6$ \citep{2012MNRAS.427.3435A, Desjacques:2016bnm,2017MNRAS.470.2617A}. As this analysis is restricted to low redshift, so we have neglected the contribution from weak lensing, which is primarily important for sources at high redshift \citep{Holz:1997ic,2010PhRvD..81l4046H, Mukherjee:2019wcg}. Though we do have not included \texttt{Euclid} in the mock catalog study,  combining all the catalogs is going to be partly useful. 

For the mock samples of GW sources, we obtain the GW spatial distribution from the same realization of the mock galaxy catalog with the population and source properties of the GW sources mentioned in Sec. \ref{GWcatalog}. Due to the larger sky coverage of \texttt{SPHEREx} ($f_{sky}= 0.65$) than \texttt{DESI} ($f_{sky}= 0.3$),  the total number of GW sources which will be overlapping with \texttt{DESI} sky area is going to be less than the total number of sources overlapping with \texttt{SPHEREx}. As a result, we consider we obtain the GW samples for the corresponding values of $N_{GW}$ shown in Fig. \ref{fig:plot_powerlaw}. The overlap for this with GW merger rate with different spectroscopic galaxy surveys is shown in Fig. \ref{fig:overlap} which shows the possible synergies expected between \texttt{LVK}+\texttt{DESI} and \texttt{LVK}+\texttt{SPHEREx}.  {The typical relative luminosity distance error is around  $20-30\%$ for the mock samples having a matched filtering SNR greater than ten for a network of four detectors. This leads to the maximum value of the Fourier modes parallel to the line of sight $k^l_{max} < 0.02$ h/Mpc. Similarly, the typical sky localization error of the GW sources which are above at least a matched filtering SNR of $10$ exhibits a sky localization error greater than around $\Delta \Omega_{GW}= 25$ sq. deg, which cast comoving Fourier modes $k^p_{max}<0.1$ h/Mpc for nearly all the GW mock samples considered in this analysis. These scales are below the quasi non-linear scale and hence a diagonal covariance matrix we consider in this analysis remains valid \citep{Blot:2015cvj,Howlett:2017vwp,Sugiyama:2019ike,Blot:2020cbi,Chartier:2021frd}. For the implementation of this technique on data, we need to consider the covariance matrix from mocks associated with individual missions such as \texttt{DESI} and \texttt{SPHEREx}.}

\section{Forecast for the cosmological parameters}\label{results}
We explore the synergy between \texttt{LVK}+\texttt{DESI} and \texttt{LVK}+\texttt{SPHEREx}, by performing a joint Bayesian analysis following the framework developed in \cite{Mukherjee:2020hyn,Mukherjee:2020mha}. The posterior on the cosmological parameters can be inferred using the gravitational wave data vector $\vec{\vartheta}_{GW} \equiv \{ d^i_l,\, \theta^i_{GW},\, \phi^i_{GW}\}$ and the galaxy data vector $\vec d_{g} \equiv \{\delta_g( z_g^i,\,\theta^i_{g},\, \phi^i_{g})\}$ by combining Eq. \eqref{eqbayes1} and Eq. \eqref{eqbayes5} as
\begin{align}\label{posterior-1}
   \mathcal{P}(\Theta_c|\vec{\vartheta}_{GW}, \vec d_{g})&\propto  \iint d\Theta_n\, dz \, \prod_{i=1}^{N_{GW}}\, \, \mathcal{L}(\vec{\vartheta}_{GW}| \{P_{gg}(z)\}, \Theta_n, \vec d_g(z))\nonumber \\ & \times \mathcal{P}(\vec d_g| \{P_{gg}(z)\}) 
   \mathcal{P}({\{d^i_l\}}_{GW}|z, \Theta_c, \{\hat \theta^i,\, \hat \phi^i\}_{GW})\nonumber \\& \times\, \Pi(z) \Pi(\Theta_n)\Pi(\Theta_c).
\end{align}

For the \texttt{DESI} survey, we take the number density of BGS  and ELGs  according to the redshift distribution shown in \citet{desi}. For \texttt{SPHEREx}, we only consider sources with the corresponding number distribution which are measurable with the redshift error $\sigma_z = 0.003 (1+z)$ \citep{spherex}. We take priors as $\Pi(H_0)= \mathcal{U}[20, 150]$ km/s/Mpc, $\Pi({w_0})= \mathcal{U}[-1.5, -0.3]$, $\Pi(b_{GW})=  \mathcal{U}[0, 6]$, and $\alpha=\mathcal{U}[-4, 4]$ for both \texttt{LVK}+\texttt{DESI} and \texttt{LVK}+\texttt{SPHEREx} analysis. For the implementation of the cross-correlation technique, we have chosen a bin width $\Delta z=0.1$, and our the final posteriors do not any significant departure for change in the bin choices to $\Delta z=0.05$.

\subsection{Measurement of the Hubble parameter and GW bias parameter}
Using the cross-correlation technique, we investigate the measurability of the Hubble constant from the combination of \texttt{LVK}+\texttt{DESI} and \texttt{LVK}+\texttt{SPHEREx} missions. The sky-fraction of \texttt{DESI} is smaller than the sky-fraction available from \texttt{SPHEREx}. So, the number of GW sources overlapping with the \texttt{DESI} survey will be less than the \texttt{SPHEREx} survey. By combining the 5-years observational run of \texttt{LVK}, we show the forecast for the measurement of the Hubble constant, after marginalizing over the GW bias parameter $b_{GW}$ and its redshift dependence $\alpha$ for the Lambda-Cold Dark Matter model of cosmology in Fig. \ref{fig:h0} for \texttt{LVK}+\texttt{DESI} (in cyan) and \texttt{LVK}+\texttt{SPHEREx} (in red). The constraints are going to be about $\sim 2\%$ and $\sim 1.5\%$ measurement of the value of Hubble constant from \texttt{LVK}+\texttt{DESI} and \texttt{LVK}+\texttt{SPHEREx} respectively. The difference in the measurability is mainly due to the smaller sky-fraction of \texttt{DESI} survey than \texttt{SPHEREx}. Also, as shown previously in \citet{Mukherjee:2020hyn, Mukherjee:2020mha} this method will also be able to measure for the first time GW bias parameter $b_{GW}(z)= b_{GW}(1+z)^\alpha$ and shed light on its redshift dependence from both \texttt{LVK}+\texttt{DESI} and \texttt{LVK}+\texttt{SPHEREx}. With the availability of more GW sources, the error on the cosmological parameters will scale as $N_{GW}^{-1/2}$ \citep{Mukherjee:2020hyn, Mukherjee:2020mha}. This method can take care of the peculiar velocity correction by implementing the method proposed by \citet{Mukherjee:2019qmm} on \texttt{DESI} and \texttt{SPHEREx} data.  When this method will be applied to the data, we will consider the covariance matrix estimated from simulations including the possible systematic specifics to these surveys. 

\subsection{Measurement of dark energy equation of state}
The dark energy equation of state $w_0$ is another interesting quantity to understand the nature of cosmic acceleration. We have explored the measurability of this parameter from \texttt{LVK}+\texttt{DESI} and \texttt{LVK}+\texttt{SPHEREx} combinations after marginalizing over the GW bias parameters and show the corresponding 1-D marginalized posterior in Fig. \ref{fig:w0}. This shows that, we can make about a $10\%$ measurement from \texttt{LVK}+\texttt{DESI} and a $\sim8\%$ measurement from \texttt{LVK}+\texttt{SPHEREx} within the five years of observation.  Joint estimation of the multiple cosmological parameters is possible, but the measurability will significantly degrade. As a result, to get interesting constraints within the first five years of \texttt{LVK} observation, it is appropriate to get constraints only on $w_0$, and in the future with more sources, one can measure the parameters jointly. The difference in the peak value of the posterior distribution and the injected mean value is due to less number of sources, and with more sources, unbiased estimation is possible. With a longer observation time and more nearby GW objects sources, it will be possible to obtain better constraints on the cosmological parameters \citep{Mukherjee:2020hyn, Mukherjee:2020mha}. For the third generation GW detectors such as \texttt{Einstein Telescope} \citep{Punturo:2010zz}, and \texttt{Cosmic Explorer} \citep{Reitze:2019iox}, with a large number of sources, this method will be able to reach much better precision in the measurement of $w_0$. This method will also make it possible to measure the redshift dependence of the dark energy equation of state ($w(z)= w_z(z/(1+z))$) in a longer time scale with more GW sources \citep{Mukherjee:2020hyn}.

\section{Conclusion}\label{conclusion}
We explore the synergies between the GW detectors such as \texttt{LIGO}, \texttt{Virgo}, and \texttt{KAGRA} and the near-term spectroscopic surveys such as \texttt{DESI}, \texttt{SPHEREx}, and \texttt{EUCLID} to measure the clustering redshift for the dark standard sirens. One of the key requirements to infer the clustering redshift is the overlap in the redshift and sky-position of the GW sources with the spectroscopic galaxies. The redshift overlap between the detected GW sources and spectroscopic galaxy surveys depends on a few factors such as the merger rate of the GW sources, the horizon for a network of GW detectors, and the redshift space window function of spectroscopic surveys. We show in Fig. \ref{fig:overlap} that for GW mergers rates motivated by the \citet{Madau2014} SFR for a delay time $t_d^{eff}< 2$ Gyr, one would expect overlapping GW sources from the redshift range which are accessible from the combination of the BGS and ELG from \texttt{DESI} \citep{desi} having a redshift error $\sigma_z= 0.001(1+z)$, \texttt{SPHEREx-A} and \texttt{SPHEREx-B} galaxy surveys with redshift errors $\sigma_z = 0.003(1+z)$ and $\sigma_z = 0.01(1+z)$ respectively \citep{spherex}, and \texttt{Euclid} \citep{Euclid:2019clj} have $\sigma_z = 0.001(1+z)$. However, due to detector horizon up to a redshift $z\sim 1.1$, sources beyond this redshift do not contribute to the estimation of the cosmological parameters. 

For the implementation of the cross-correlation study and exploring a large parameter space of the GW sources, we need to combine all the different spectroscopic surveys mentioned above. In this work we focus only on \texttt{DESI} and \texttt{SPHEREx-A}, however some more improvements can also be made if we can combine \texttt{Vera Rubin Observatory} \citep{2009ApJ...695..287R} and \texttt{Roman Telescope} \citep{2012arXiv1208.4012G, 2013arXiv1305.5425S, Dore:2018smn}. We will explore the synergy with it in future work (under preparation).

To explore the measurability of the cosmological parameters using the cross-correlation technique, we simulate GW sources up to redshift $z=0.5$ using \texttt{IMRPhenomPv2} waveform model. We consider five years of observation time with a $50\%$ duty cycle. We perform a GW source parameter estimation in a Bayesian framework using the package \texttt{Bilby} \citep{bilby}. We perform a Bayesian parameter estimation of the GW sources to correctly include degeneracy between different parameters and also include the non-Gaussian posterior distributions of the source parameters. We restrict our analysis only to the low-redshift mock samples to reduce the computational cost in estimating the GW source parameters. Also, GW sources which are at high redshift, the error in the luminosity distance and sky localization are poor and hence will not make a significant difference in the measurability of the cosmological parameters. However, in a future work (under preparation) we will use approximate-likelihood to explore the synergy with \texttt{Euclid}, \texttt{Vera Rubin Observatory}, and \texttt{Roman Telescope}. 

We find that both the combinations \texttt{LVK}+\texttt{DESI} and \texttt{LVK}+\texttt{SPHEREx} are very powerful in exploring the clustering redshift of the dark standard sirens and can be used to measure the expansion history and GW bias parameters. We find that \texttt{LVK}+\texttt{DESI} and \texttt{LVK}+\texttt{SPHEREx} can make about $\sim 2\%$ and $\sim 1.5\%$ measurement of the value of Hubble constant $H_0=67.7$ km/s/Mpc in five years of observation of the \texttt{LVK} network of detectors with $50\%$ duty cycle. This measurement is after marginalization of the selection effect that is captured by the GW bias parameters. In this setup, we can also explore the redshift dependence of the GW bias parameter which captures how the GW sources trace the underlying dark matter distribution. Using the cross-correlation technique, we can also measure the value of dark energy equation of state $w_0$ (keeping $H_0$ and $\Omega_m$ fixed) with about $10\%$ and $8\%$ precision from  \texttt{LVK}+\texttt{DESI} and \texttt{LVK}+\texttt{SPHEREx} respectively. The primary reason for the difference in the performance between \texttt{LVK}+\texttt{DESI} and \texttt{LVK}+\texttt{SPHEREx} is due to larger sky-fraction available to \texttt{SPHEREx} than \texttt{DESI}. In the future, with the operation of \texttt{LIGO-India}, a further improvement in the cross-correlation study is possible due to improvement in the sky-location error  \cite{Unnikrishnan:2013qwa}. To achieve both accurate and precise measurement of expansion history, it is also important to mitigate the possible systematics which can arise from the GW waveform modeling and instrument calibration \citep{Sun:2020wke, Bhattacharjee:2020yxe, Vitale:2020gvb}. Though in this work, we mainly explore the synergies with the galaxy surveys and \texttt{LVK} detectors to explore cosmology, in a future analysis impact of waveform systematics and instrument calibration will be explored. In addition to the dark sirens, detection of bright \textit{standard sirens} \citep{Chen:2017rfc,Feeney:2018mkj, Mukherjee:2020kki} can provide additional improvement in measuring the Hubble constant, dark energy equation of state. 

In summary, this study shows the possible synergies between the GW detectors and galaxy surveys in exploring cosmology. This multi-messenger avenue will shed light not only in resolving the tension in the value of Hubble constant between the low redshift probes \citep{Efstathiou:2013via,Riess:2019cxk,Riess:2019qba,Wong:2019kwg,Freedman:2019jwv,Freedman:2020dne,Soltis:2020gpl} and the high redshift probes \citep{Spergel:2003cb, Spergel:2006hy,2011ApJS..192...18K, Ade:2013zuv, Ade:2015xua,Alam:2016hwk, Aghanim:2018eyx,Verde:2019ivm, eBOSS:2020yzd,ACT:2020gnv, DiValentino:2021izs}, but will also provide an independent measurement of the dark equation of state which can bring a better understanding of its nature and its interaction with the tensor sector. This synergistic study will provide the first measurement of the GW bias parameter and its redshift dependence and will shed light on its connection with the dark matter distribution in the Universe. Finally, the cross-correlation technique between the GW sector and galaxies will be able to test the general theory of relativity which is currently not feasible from only the electromagnetic observations \citep{Mukherjee:2019wfw, Mukherjee:2019wcg, Mukherjee:2020mha}. 

\section*{Acknowledgements}
The authors are grateful to Archisman Ghosh for reviewing the manuscript as a part of LVK publication and presentation policy. SM acknowledges useful inputs from Martin Hendry on this work. A part of this work is carried out under the Master's program at the University of Amsterdam. This work is part of the Delta ITP consortium, a program of the Netherlands Organisation for Scientific Research (NWO) that is funded by the Dutch Ministry of Education, Culture, and Science (OCW). S.M. is supported by the Simons Foundation. Research at Perimeter Institute is supported in part by the Government of Canada through the Department of Innovation, Science and Economic Development and by the Province of Ontario through the Ministry of Colleges and Universities. This analysis is carried out at the Infinity cluster, the Horizon cluster hosted by Institut d'Astrophysique de Paris, and the Lisa cluster hosted by the University of Amsterdam. We thank Stephane Rouberol for smoothly running the Infinity cluster and the Horizon cluster. We acknowledge the use of following packages in this work: \texttt{Astropy} \citep{2013A&A...558A..33A,2018AJ....156..123A}, \texttt{Bilby} \citep{bilby}, \texttt{DYNESTY} \citep{2020MNRAS.493.3132S}, \texttt{emcee: The MCMC Hammer} \citep{2013PASP..125..306F}, \texttt{Giant-Triangle-Confusogram} \citep{Bocquet2016},  \texttt{IPython} \citep{PER-GRA:2007},  \texttt{Matplotlib} \citep{Hunter:2007},  \texttt{nbodykit} \citep{Hand:2017pqn},  \texttt{NumPy} \citep{2011CSE....13b..22V}, and  \texttt{SciPy} \citep{scipy}. The authors thank the  \texttt{SPHEREx} science team for providing the galaxy distribution. The authors would like to thank the  LIGO/Virgo scientific collaboration for providing the noise curves. LIGO is funded by the U.S. National Science Foundation. Virgo is funded by the French Centre National de Recherche Scientifique (CNRS), the Italian Istituto Nazionale della Fisica Nucleare (INFN), and the Dutch Nikhef, with contributions by Polish and Hungarian institutes. This material is based upon work supported by NSF’s LIGO Laboratory which is a major facility fully funded by the National Science Foundation.

 \section*{Data Availability}
The data underlying this article will be shared at reasonable request to the corresponding author. 

\bibliographystyle{mnras}
\bibliography{main}
\label{lastpage}
\end{document}